# tRNA – alike in *Nanoarchaeum equitans* ?

Bibekanand Mallick [a], Jayprokas Chakrabarti [a, b,*],

Zhumur Ghosh [a], Smarajit Das [a] and Satyabrata Sahoo [a]

[a] Computational Biology Group(CBG)

Theory Department

Indian Association for the Cultivation of Science

Calcutta 700 032 INDIA

[b] Biogyan

BF 286 , Salt Lake

Calcutta 700 064 INDIA

[*] Corresponding author

Email: tpjc@iacs.res.in

Also: biogyan@vsnl.net

Fax: 91-33-24732805

Telephone: 91-33-24734971 ext. 307

The recent algorithm[1] for five split tDNAs in *N.equitans* is new. It locates missing tRNA$^{Trp}$, tRNA$^{iMet}$, tRNA$^{Glu}$ and tRNA$^{His}$. But the split tRNA$^{Trp}$(CCA) solution is anomalous ; the tRNA$^{iMet}$ solution[1] lacks cognition elements for aminoacylation. In view therefore we present here alternate non-split composite solutions for tRNA$^{Trp}$, tRNA$^{iMet}$, tRNA$^{Glu}$ and tRNA$^{His}$.

Earlier[2] tRNA genes in *N.equitans* were exhaustively explored. The remarkable algorithms tRNAScan-SE[3] and ARAGORN[4] located all tRNAs except tRNA$^{Trp}$, tRNA$^{iMet}$, tRNA$^{Glu}$ and tRNA$^{His}$. The new algorithm of Randau et al[1] locates these missing ones.

However, the tRNA$^{Trp}$(CCA) reported[1] is anomalous : 1) There is GG preceding the anticodon. We studied all archaeal tRNA$^{Trp}$(CCA) and found this to be an exception. U33 is known[5] to contribute to tRNA-ribosomal binding. Its absence is puzzling. 2) Further archaeal tRNA$^{Trp}$(CCA) always have discriminator base A73. This discriminator A73 is of modest preference for aminoacylation[6]. Randau et al[1] have C73.

Again, the 73$^{rd}$ discriminator base of archaeal tRNA$^{iMet}$(CAU) is always A73. But, tRNA$^{iMet}$(CAU) solution[1] is anomalous, it has U73 .

In the absence of conclusive aminoacylation experiment and the anomalies listed above, we reanalyzed the missing tRNAs for Nanoarchaea. In the split-tRNA hypothesis[1] the structures ( 5-primed end split at 37 followed by invert-repeat element , 3-primed end preceeded by invert-repeat element etc ) of tDNA-Glu/His are similar to tDNA-Trp/iMet. If tRNA-Trp/iMet are anomalous, how functional are tRNA-Glu/His? Are there other solutions? From the classic work[7] (and the references therein) on tRNA it is known that archaeal tRNA harbour noncanonical introns. Canonical introns are located between bases 37 and 38 of tRNA; noncanonical introns occur elsewhere. We looked for the possibility that tDNA-Trp/iMet/Glu/His have noncanonical introns. We found composite solutions that do not suffer from the anomalies above. These solutions are:

tRNA<sup>Trp</sup> gene-151992-152078
5'-TAGAAAAATT**TTTAAA**TATCTATCTATTGCAATCTC**GGGGCCGTAGCTCAGCCAGGCAGAGCGGCGG**ATTCGAAGCCGAAG**CTCCAGACCCGTAGGTCGGGGGTTCGAATCCCCCCGGCCCCA** -3'

tRNA<sup>iMet</sup> gene-35249-35429
5' TCGTTAATTCTTACAGTAACATT**TATAAA**TGGTTTTGTTATAACTTACTA**CGCGGGGTGGGGCAGCCTGGAGTGCCCCGGGGGCTCATATCCCCCTGGCCGCCTTTTTTTCATATTTAATGGACCCGCCGG**ATTCGAACCCGGGGCCTCCGCCTTGCGAGGGCGGCGT**CCTACCGCTGGA**CTACGGGCCCGGTTTCG**ATTTTAGATACAAAATAAATACATTTTT**TGTAA-3'

tRNA<sup>Glu</sup>(CTC) gene-151992-152078
5'-AATT**TTTAAA**TATCTATCTATTGCAATCTC**GGGGCCGTAGCTCAGCCAGGCAGAGCGGCGG**ATTCGAAGCCGAAG**CTCCAGACCCGTAGGTCGGGGGTTCGAATCCCCCCGGCCCCA**-3'

**tRNA^His(ATG) gene-327362-327525**

5'-ATAATT`TTTAAA`TCGTTTCTTTTATTCTATTG`GCCGCCGTAGCTCAGC`
`GGTCAGAGCGCCCGGC`TCATAGCATGGGC`TATGAGCTCTGACCCGA`
`AAGGGGATGATCTCGGG`GGCTCTTATGCCCCCTCGTGAGAA`ACCGG`
`GAGGTCGCGGGTTCGAATCCCGCCGGCGG`CATCACAATTTTTATATA
AACCTAACC-3'

Here we have marked tRNAs in red, introns in blue, the conserved archaeal Box A promoter-elements[8] in green. We found the right secondary structures for all these tRNAs, and the bulge-helix-bulge (BHB) motifs. Note, for instance, the following important features of this tRNA^Trp(CCA): U8, A14, A21, U33, G18:U55, G19:C56, U54:A58 and G30:C40, the anticodon CCA at 34,35,36, and finally A73. These bases/base-pairs are conserved in all tRNA^Trp(CCA) in archaea. tRNAScan-SE identifies bases 151992 to 152081 as tRNA^Ser(CGA). Note there is another tRNA^Ser(CGA) between 486337 and 486426. The one between 151992 and 152081 is unlikely to be tRNA^Ser(CGA): none of the conserved bases/base pairs of archaeal tRNA^Ser viz. G1:C72, G18:U55, G19:C56, U54:A58, G26:U44, G53:U61, U33, G73 appear. Again, the Variable-arm is absent. It is known[9] G73 and Variable-arm contain identity elements for Ser-RS.

From our study of 22 fully sequenced archaea, the 73rd discriminator base of tRNA^iMet(CAU) is A73. Our tRNA^iMet(CAU) has A73. It shares all features of archaeal tRNA^iMet(CAU).

Remarkably our tRNA<sup>Glu</sup>(CUC) and tRNA<sup>Trp</sup>(CCA) overlap with one another. Note the tDNA<sup>Glu</sup>(CUC) has a noncanonical intron at 33. tDNA<sup>Trp</sup>(CCA) has a noncanonical intron at 30. *N.equitans* has the smallest genome known. Noncanonical introns here compactify two tDNAs. Interestingly, this compactification is at work for tRNA<sup>His</sup> as well.

Codon usage study of histidine in 22 archaea reveals the ratio of the number of CAU-codon to CAC-codon to be anomalously high in *N.equitans*. Amongst archaea *N.equitans* is special in this respect. For tRNA<sup>His</sup> ATG is the likely anticodon. This is precisely what we found: tDNA<sup>His</sup>(ATG) lying between 327362 and 327520. It has two noncanonical introns located between 32/33 and 71/72 of 13 and 25 bases respectively. In addition to these, there is the canonical intron of 53 bases. Remarkably again, this tDNA<sup>His</sup>(ATG) overlaps with tDNA<sup>eMet</sup>(CAU), located between 327362 and 327500. tDNA<sup>eMet</sup>(CAT) has a canonical intron of 66 bases.

Randau et al's split-tRNA solutions are new. Splitting decompactifies the genome. Further some of the split solutions are anomalous. Our solutions have overlapping composite tRNA genes[10]. tRNA genes are woven together by introns. They appear just suited for *N.equitans* that has the smallest genome.

**Reference:**

1. Randau, L., Munch,R, Hohn, M.J., Jahn,D and Söll,D. *Nature* **433**, 537–541 (2005).

2. Waters, E. *et al. Proc. Natl Acad. Sci.* USA **100**, 12984–12988 (2003).

3. Lowe, T. M. and Eddy, S. R. *Nucleic Acids Res.* **25**, 955–964 (1997).

4. Laslett, D., and Canback, B. *Nucleic Acids Res.* **32**, 11-16 (2004).

5. Ashraf, S.S. *et al. RNA* **5**, 503-511(1999).

6. Guo,Q. *et al. J.Biol.Chem.* **277**(16), 14343-14349 (2002).

7. Marck, C and Grosjean,H. *RNA* **9**, 1516-1531(2003).

8. Palmer J R and Daniels C J. *Journal of Bacteriology* **177**(7), 1844-1849 (1995).

9. Giege, R., Sissler, M., and Florentz, C. *Nucleic Acids Res.* **26**, 5017-5035 (1998).

10. Reichert,A., Rothbauer,U. and Mörl, M. *The Journal of Biological chemistry.* **273**(48), 31977–31984(1998).